\documentclass[pre,twocolumn,aps,eqsecnum]{revtex4}
\usepackage{epsfig}
\usepackage{graphicx}
\usepackage{latexsym}

\begin{document}

\title{Shear-transformation-zone theory of plastic deformation near the glass transition}

\author{ J. S. Langer}
\affiliation{Dept. of Physics, University of California, Santa Barbara, CA  93106-9530}

\date{\today}

\begin{abstract}
The shear-transformation-zone (STZ) theory of plastic deformation in glass-forming materials is reformulated in light of recent progress in understanding the roles played the effective disorder temperature and entropy flow in nonequilibrium situations. A distinction between fast and slow internal state variables  reduces the theory to just two coupled equations of motion, one describing the plastic response to applied stresses, and the other the dynamics of the effective temperature.  The analysis leading to these equations contains, as a byproduct, a  fundamental reinterpretation of the dynamic yield stress in amorphous materials. In order to put all these concepts together in a realistic context, the paper concludes with  a reexamination of the experimentally observed rheological behavior of a bulk metallic glass. That reexamination serves as a test of the STZ dynamics, confirming that system parameters obtained from steady-state properties such as the viscosity can be used to predict transient behaviors.  
\end{abstract}
\maketitle

\section{Introduction}
\label{intro}

For over a decade, my coworkers and I have been developing a shear-transformation-zone (STZ) theory of plastic deformation in noncrystalline solids. \cite{FL98,FALK99,LP03,JSL04,PECHENIK05,BLP07I,BLP07II} Our goal has been to construct a phenomenological description of amorphous plasticity based on physical principles and molecular models, and yet simple enough to be useful for predicting the performance of real materials.  At the molecular level, amorphous solids are structurally no more complicated than fluids.  They do, of course, exhibit highly non-fluidlike properties -- rigidity, jamming, and the like.  Nevertheless, their underlying simplicity implies that their behaviors might exhibit some degree of universality.  My purpose in this paper is to move the STZ theory further toward  such a description. In particular, I propose to start with a few basic features of a flow-defect theory of amorphous plasticity and then to see how far it is possible to go using primarily symmetry, conservation laws, and thermodynamics.  

From its inception \cite{FL98}, the STZ theory was intended to be an extension of the flow-defect theories of Turnbull, Cohen, Spaepen, Argon and others \cite{TURNBULL-COHEN70,SPAEPEN77,ARGON79,SPAEPEN81,ARGON83} in which localized deformable clusters of molecules allow noncrystalline solids to undergo irreversible shear strains in response to applied stresses. The reformulation of the STZ theory presented here is motivated primarily by the emergence of the effective disorder temperature as a key internal state variable in nonequilibrium theories of amorphous plasticity.  The dynamic role of the effective temperature has been explored in remarkable detail recently by Haxton and Liu \cite{HAXTON-LIU07} in their extensive molecular-dynamics simulations of a simple, two-dimensional glass forming material.  Much of the content of the present paper was developed during the attempt by Manning and myself \cite{JSL-MANNING07} to interpret the Haxton-Liu data.

The relation between the population of STZ's (or flow defects) and an intensive variable such as the effective temperature has a long history.  Earlier investigators, notably Cohen and Turnbull \cite{COHEN-TURNBULL59} and Spaepen \cite{SPAEPEN77}, described the intrinsically disordered state of noncrystalline materials by a free volume $v_f$.  Those authors perceptively recognized that the relevant definition of $v_f$ is not as an extensive excess volume measured from some densely packed state, but as an intensive quantity -- the inverse of the derivative of the configurational entropy (i.e. the entropy associated with molecular configurations, without kinetic contributions) with respect to the volume.  Thus they proposed that the density of flow defects be proportional to a Boltzmann-like factor, $\exp\,(- {\rm constant}/v_f)$, and not just to $v_f$ itself.  

In \cite{JSL04}, I argued that the appropriate generalization of free volume in plasticity theory is an effective temperature $T_{e\!f\!f}$ that characterizes the state of configurational disorder in the system.  In analogy to $v_f$, $T_{e\!f\!f}$ is the inverse of the derivative of the configurational entropy with respect to configurational energy.  $T_{e\!f\!f}$ equilibrates to the ambient temperature $T$ at high $T$, but may fall out of equilibrium at low $T$ where disorder is generated by the molecular rearrangements that accompany mechanical deformation.  Throughout this paper, as in \cite{JSL04}, I define $T_{e\!f\!f}= T_Z\,\chi$, where $E_Z = k_B\,T_Z$ is a characteristic STZ formation energy, so that the STZ density is proportional to $\exp\,(-1/\chi)$. This is a direct analog of the free-volume formula and, in fact, reduces to it in the case of a system under constant pressure with a positive ``effective'' thermal expansion coefficient.  

With this definition of $\chi$, the ``bottom line'' of the analysis to be presented here is that the STZ theory -- in most but not all circumstances -- can be reduced to an expression for the plastic part of the rate-of-deformation tensor, 
\begin{equation}
\label{D-chi-s}
D_{ij}^{pl}= e^{-1/\chi}\,f_{ij}({\bf s},T),
\end{equation}
supplemented by an equation of motion for $\chi$, which appears here only in the Boltzmann-like prefactor. In Eq.(\ref{D-chi-s}), ${\bf s}$ is the deviatoric stress tensor and $T$ is the bath temperature.  The tensor function $f_{ij}({\bf s},T)$ contains all the dynamical details pertaining to STZ transitions; the prefactor $\exp\,(-1/\chi)$ determines the density of STZ's.  This clean separation between the ${\bf s}$ and $\chi$ dependence of the plastic strain rate was emphasized  by Shi et al. \cite{SHIetal07} as a characteristic feature of the STZ theory. It is derived here in Section \ref{STZequations}; and the equation of motion for $\chi$ is derived in Section \ref{Teff}. 

This paper starts, in Sec.\ref{STZbasics}, with a reexamination of the basic assumptions upon which the STZ theory is constructed.  The STZ equations of motion are derived in Section \ref{STZequations}, and a  general expression for the yield stress is derived and interpreted in Section \ref{yieldstress}.  The latter result brings new physical insight to the choice of material-specific quantities such as the STZ transition rates.  Sections \ref{viscosity} and \ref{Teff}, respectively, contain a derivation of the STZ formula for the Newtonian viscosity and a discussion of effective-temperature dynamics. As a way of putting all these concepts together in a realistic context, the main part of the paper concludes in Sec. \ref{metglass} with a reexamination of the extensive set of rheological measurements of a bulk metallic glass by Lu et. al., \cite{LUetal03}. That reexamination, a reworking of \cite{JSL04} and \cite{FLP04}, serves as a test of the STZ dynamics, confirming that system parameters obtained from steady-state properties such as the viscosity can be used to predict transient behaviors. Finally, Section \ref{conclusions} contains a few brief remarks about outstanding problems. 

\section{Basic assumptions of the STZ theory}
\label{STZbasics}

The starting point for all of the analysis in this paper is a dynamic model of a noncrystalline material in which a disordered arrangement of molecules (or bubbles, or sand grains, etc.) interact with each other via short ranged forces.  This system is subject to driving forces that cause it to deform and, in some circumstances, to flow continuously.  It is dynamic in the sense that it possesses an intrinsic time scale, say $\tau_0$, that characterizes the rate at which it responds to microscopic perturbations.  In molecular materials, $\tau_0$ is an internal vibration period of order femtoseconds.  The theoretical challenge is to understand the mechanism by which these fast, molecular interactions generate viscous responses on time scales of order seconds or longer. 

To construct a theory of amorphous plasticity for such a model, I have found it useful to watch computer-generated, moving pictures of its behavior as it undergoes slow, steady, shear deformation. These pictures can be found on the websites of the authors of Refs. \cite{HAXTON-LIU07} and \cite{DENNIN-OHERN07}. Their most striking feature is that the irreversible molecular rerrangements are sporadic, short lived, spatially isolated events.  These are the STZ transitions.  They occur within a persistently noisy environment in which thermally and mechanically generated fluctuations bring groups of molecules into and out of configurations in which they can undergo irreversible shear transformations. That is, the fluctuations create and annihilate STZ's.  At low temperatures, below the glass transition, rearrangements are driven primarily by the applied stress; and an STZ, once formed, rapidly undergoes a shear transition if it is aligned favorably with respect to the stress. At higher temperatures and small driving forces, i.e. in the viscous regime above the glass transition, STZ transitions are thermally driven and the applied stress simply biases them the direction of the average plastic shear rate.  In either limit, the STZ transitions are rare inelastic events occurring within an otherwise solidlike elastic material.  

It is especially easy to observe STZ rearrangements in two-dimensional models, where their cores are T1 events in which two nearest-neighbor molecules move away from each other and two next-nearest neighbors come together.  No two STZ transitions ever seem to look exactly the same.  The T1 events at their cores occur at varying angles with respect to the shear direction; and varying numbers of nearby molecules participate in the overall motions.  This statistical variability of the STZ's is clearly due to the fluctuating environment in which they are being created and annihilated.   

Experimental observations of memory effects in simple amorphous materials tell us that the STZ's cannot be structureless objects.  The Bauschinger effect is one example where the system remembers the direction in which it has been deformed, and responds differently -- more compliantly or less so -- to further loading in different directions.  The natural way to include such effects in the theory is to let the STZ's possess internal degrees of freedom that carry information from one event to the next.  The simplest such possibility is to assume that, during their lifetimes, the STZ's are dynamic, anisotropic objects whose populations and orientations are determined by the loading history.

Once a shear transformation occurs in some direction, the molecules that composed the STZ resist further shear in the original direction but may be especially susceptible to a reverse shear.  That is, the first transition redistributes the local stresses in such a way as to favor a reverse transition if the stress changes sign.  There is no strong requirement that the reverse transition bring the molecules back to exactly their original positions; but it is this approximate picture that suggests a two-state model of STZ's.  The two-state behavior has been observed directly by Lundberg et al. \cite{DENNIN-OHERN07} in parallel bubble-raft experiments and two-dimensional foam simulations.  These authors find that a T1 event is  often mechanically reversible (although still dissipative) if the direction of the applied shear is reversed shortly after the event occurs.  On the other hand, if the shear is not reversed until after other events have occurred nearby, then the memory is lost, i.e. the STZ is annihilated.
 
Thus, the two-state STZ's are the memory carriers in amorphous materials.  If they are randomly oriented and present in substantial numbers, the material is deformable in all directions because there are STZ's available to respond to all directions of applied stress.  If they are rare, deformation is difficult in any direction.  If they are all oriented in the direction of a shear stress, then further deformation in that direction is impossible -- the material is jammed -- but response to a reverse stress is relatively easy.  Annihilation of existing STZ's and creation of new ones without orientational bias are the mechanisms by which memory is lost in these systems.  

During its lifetime, each STZ has two preferred spatial orientations.  One of these orientations, say the ``$+$'' state, will be closest to the direction of the deviatoric stress tensor ${\bf s}$ and therefore most stable.  In two dimensions, the other STZ state, say ``$-$'', is perpendicular to the first and is therefore least stable with respect to ${\bf s}$ but most stable with respect to $-{\bf s}$. The situation is more complicated in three dimensions where the two STZ states need not be symmetrically oriented with respect to ${\bf s}$; but it must still be true that, except in very special cases, if we are given a stress tensor ${\bf s}$ and a pair of STZ orientations, we can identify the ``$\pm$'' orientations unambiguously.  

Let the symbol $\omega$ denote the orientation of an STZ, that is, the orientation of the axes along which its $\pm$ states are defined, and let the symbol $\ell$ denote other characteristics of the STZ such as its size or its transition threshold.  Then let $n_{\pm}(\omega,\ell)$ be the associated STZ density.  In accord with the preceding discussion, a  master equation for this set of these densities has the form
\begin{eqnarray}
\label{ndot}
\nonumber
&&\tau_0\,\dot n_{\pm}(\omega,\ell) ={\cal R}_{\ell}(\pm s_{\omega})\,n_{\mp}(\omega,\ell)-{\cal R}_{\ell}(\mp s_{\omega})\,\,n_{\pm}(\omega,\ell)\cr\\&&+ 
 \bigl[\Gamma({\bf s}) + \rho(T)\bigr]\left[{n_{\infty}\over 2}\,e^{-\theta_{\ell}/\chi} - n_{\pm}(\omega,\ell)\right],~~~~~~~~~
\end{eqnarray} 
where $\tau_0$ is the molecular time scale mentioned previously.  The right-hand side of Eq.(\ref{ndot}) consists of two  parts, one that couples the STZ's to the stress in a way that conserves the total number of STZ's and the configurational entropy, and a second that describes the rate at which STZ's are created and annihilated and therefore governs the entropy flow in the system. 

The first two, entropy conserving terms on the right-hand side of Eq.(\ref{ndot}) describe the rates at which the STZ's transform between their ``$+$'' and ``$-$'' states.  ${\cal R}_{\ell}(s_{\omega})$ is the rate of transitions from ``$-$'' to ``$+$'', and $s_{\omega}$ is the projection of ${\bf s}$ onto the axes defined by $\omega$, .  By symmetry, the rate of reverse transitions is ${\cal R}_{\ell}(-s_{\omega})$.  

The second pair of terms on the right-hand side describes the rate of STZ creation and annihilation.  Here, $(\Gamma + \rho)/\tau_0$ is an attempt frequency, i.e. a noise strength, consisting of incoherent mechanical and thermal parts, $\Gamma({\bf s})$ and $\rho(T)$ respectively.  The exponential function $\exp\,(-\theta_{\ell}/\chi)$ is the {\it a priori} probability of occurence of a state with  energy  $E_{\ell} = k_B\,T_Z\,\theta_{\ell}$; and $n_{\infty}$ is a reference density of order an inverse molecular volume.  $\Gamma({\bf s})$ is a measure of the strength of the mechanically generated noise that accompanies plastic deformation; it vanishes when the driving force ${\bf s}$ is absent.  An explicit expression for $\Gamma({\bf s})$ will be derived in Sec. \ref{STZequations}.  

$\rho(T)$  is the super-Arrhenius part of the rate of thermally activated molecular rearrangements.  Its molecular origin lies at the heart of any theory of the glass transition, and is discussed in detail -- albeit quite speculatively -- in my paper on an ``excitation-chain theory of glass dynamics. \cite{JSL-XCPRE06,JSL-XCPRL06} $\rho(T)$ vanishes when $T$ is less than the glass transition temperature $T_0$ and is equal to unity when $T$ is above the super-Arrhenius region, say, $T > T_A$.  It is  convenient to write
\begin{equation}
\label{alphadef}
\rho(T) = \cases{e^{-\alpha(T)} & for $T>T_0$\cr 0 & for $T<T_0$}
\end{equation}
where $\alpha(T)$  vanishes for $T>T_A$ and diverges, perhaps like $(T-T_0)^{-1}$, as $T$ approaches $T_0$ from above.  When there is no external driving and the effective temperature $T_{e\!f\!f}=\chi\,T_Z$ equilibrates to $T$, the STZ creation rate is proportional to $\exp\,[- E_{\ell}/k_B\,T - \alpha(T)]$, which is the ``$\alpha$'' relaxation rate.  Importantly, the activation energy appearing here is the sum of Arrhenius and super-Arrhenius parts, $E_{\ell}$ and $k_B\,T\,\alpha(T)$ respectively.  

Two aspects of Eq.(\ref{ndot}) require extra attention.  First, note that $\Gamma({\bf s})$ and $\rho(T)$ are assumed to be independent of the STZ label $\ell$; they are noise strengths that apply equally to all molecular rearrangements.  Second, the creation and annihilation part of Eq.(\ref{ndot}) is written as a single detailed-balance relation in which the equilibrium distribution  always is proportional to the Boltzmann factor $\exp\,(- \theta_{\ell}/\chi)$, with $\chi$ rather than $T$ in the exponent.  The {\it a priori} probability of forming a configurational defect always is determined by the effective disorder temperature. 

\section{STZ Equations of Motion}
\label{STZequations}

\subsection{State Variables and Time Scales}

The next order of business is to extract as much physical insight as possible from the structure of the STZ theory summarized by Eq.(\ref{ndot}), making approximations that retain this structure, but without being specific about the model-dependent ingredients of the rate factor ${\cal R}$.  

Because only the STZ's couple to the applied stress, the plastic rate of deformation tensor $D_{ij}^{pl}$ can be written in the form 
\begin{eqnarray}
\label{Dpl1}
&&\tau_0\,D_{ij}^{pl}={\epsilon_0\over n_{\infty}}\,\int d\ell\cr&&\times\left<d_{ij}(\omega)\Bigl[{\cal R}_{\ell}(s_{\omega})\,n_-(\omega,\ell)-{\cal R}_{\ell}(- s_{\omega})\,n_+(\omega,\ell)\Bigr]\right>.~~~~~~
\end{eqnarray}
Here, $\epsilon_0$ is a shear increment of order unity.  The angular brackets denote an average over STZ orientations $\omega$ consistent with the ``$\pm$'' constraints (see \cite{PECHENIK05} for details).  The symbol $\int d\ell$ denotes a  weighted sum over the other STZ properties including transition thresholds. The traceless, symmetric tensor $d_{ij}(\omega)$ projects these transitions onto the $i,j$ axes.  

For two dimensional systems, Pechenik \cite{PECHENIK05} showed that 
\begin{equation}
d_{ij}(\omega) = 2\,\hat e_i(\omega)\,\hat e_j(\omega) - \delta_{ij},
\end{equation}
where $\hat {\bf e}(\omega)$ is a unit vector at an angle $\omega$ relative to a principal axis of the stress, with  $-\pi/4 < \omega < \pi/4$.  If this principal stress axis is at an angle, say $\phi$, with respect to the $x$ axis, then $s_{ij} = s\,d_{ij}(\phi)$, where $s$ is the (signed) deviatoric stress.  Throughout the following discussion, I adopt the convention that positive values of $s$ denote stresses in the direction that drives ``$-$'' to ``$+$'' transitions.  

The situation is more complicated in three dimensions where the tensor $d_{ij}(\omega)$ must contain more than just directional information.  Note, for example, that we need two stress values plus three angles to specify the deviatoric stress tensor ${\bf s}$.  Since the tensorial versions of the STZ equations to be used here must look the same in three as in two dimensions, I omit the detailed three dimensional analysis.  

In principle, we should solve Eq.(\ref{ndot}) separately for the $n_{\pm}(\omega,\ell)$ at each $\omega$ and each $\ell$, and then use those $n_{\pm}(\omega,\ell)$ to evaluate $D_{ij}^{pl}$ in Eq.(\ref{Dpl1}).  The results of such a calculation may be important in some situations. For example, the  distribution over orientations $\omega$ may be quantitatively relevant when the system is driven strongly away from equilibrium or when the orientation of the stress changes abruptly in time.  Similarly, at small stresses, the sum over STZ thresholds implied by $\int\,d\ell$ determines the extent of plastic deformation; only the STZ's with low thresholds undergo transitions before the system becomes jammed.  Such calculations, however, would be more laborious than is necessary for most practical purposes.  Note especially that only a narrow range of STZ thresholds is likely to be dynamically relevant. STZ's with anomalously low thresholds will have large formation energies and therefore be very rare; that is, they will be suppressed by the weight factor implicit in $\int\,d\ell$.  On the other hand, STZ's with high transition thresholds will not contribute appreciably to the deformation rate; they will be suppressed by the rate factor ${\cal R}_{\ell}(s_{\omega})$.  Thus the sum over $\ell$ seems likely to be dominated by STZ's with a single characteristic  formation energy that already has been denoted by $k_B\,T_Z$.  

If only a narrow range of values of $\ell$ is dynamically relevant, then the integration over $\ell$ in Eq.(\ref{Dpl1}) contributes just a numerical factor that we can assume already has been incorporated into the reference density $n_{\infty}$.  Then, dropping the variables $\ell$, rewrite Eq.(\ref{Dpl1}) in the form   
\begin{eqnarray}
\label{Dpl1a}
&&\tau_0\,D_{ij}^{pl}={\epsilon_0\over n_{\infty}}\,\Bigl<d_{ij}(\omega)~~~\cr &\times&\Bigl[{1\over 2}\,\Bigl({\cal R}(s_{\omega})-{\cal R}(- s_{\omega})\Bigr)\,\Bigl(n_+(\omega)+ n_-(\omega)\Bigr)\cr &-& {1\over 2}\,\Bigl({\cal R}(s_{\omega})+{\cal R}(- s_{\omega})\Bigr)\,\Bigl(n_+(\omega)- n_-(\omega)\Bigr)\Bigr]\Bigr>.~~~~~~
\end{eqnarray}

The total STZ density, normalized to $n_{\infty}$, is
\begin{equation}
\label{lambdadef}
\Lambda = {1\over n_{\infty}}\,\int d\omega\,\Bigl(n_+(\omega)+ n_-(\omega)\Bigr),
\end{equation}
which defines the dimensionless density $\Lambda$. The sum $n_+(\omega)+ n_-(\omega)$, as opposed to either of these terms separately, ought to be approximately independent of $\omega$, in which case we can remove this sum from inside the integration.  As a result, the first term on the right-hand side of Eq.(\ref{Dpl1a}) can be written in the form
\begin{eqnarray}
\label{RHSfirst}
&&{\epsilon_0\over 2\,n_{\infty}}\,\Bigl<d_{ij}(\omega)\Bigl[\Bigl({\cal R}(s_{\omega})-{\cal R}(- s_{\omega})\Bigr)\Bigl(n_+(\omega)+ n_-(\omega)\Bigr)\Bigr]\Bigr>~~~~~\cr&&=
\epsilon_0\,\Lambda\,\Bigl< d_{ij}(\omega)\,\Bigl({\cal R}(s_{\omega})-{\cal R}(- s_{\omega})\Bigr)\Bigr>.
\end{eqnarray}
Similarly, we may assume that the sum ${\cal R}(s_{\omega})+{\cal R}(- s_{\omega})$ depends only weakly on the orientation and can be replaced by an $\omega$-independent function of the magnitude of the stress, specifically,
\begin{equation}
\label {Cdef}
{1\over 2}\,\Bigl[{\cal R}(s_{\omega})+{\cal R}(- s_{\omega})\Bigr]\cong {1\over 2}\, \Bigl[{\cal R}(\bar s)+{\cal R}(- \bar s)\Bigr] \equiv {\cal C}(\bar s). 
\end{equation}
where $\bar s = \sqrt{(1/2)\,s_{ij}\,s_{ij}}$ and ${\cal R}(\pm \bar s)$ denotes the value of ${\cal R}$ for a stress of magnitude $\bar s$ oriented along the ``$\pm$'' directions.  Then the second term on the right-hand side of Eq.(\ref{Dpl1a}) can be written in the form 
\begin{eqnarray}
\label{RHSsecond}
\nonumber
&&{\epsilon_0\over 2\,n_{\infty}}\,\Bigl<d_{ij}(\omega)\Bigl[\Bigl({\cal R}(s_{\omega})+{\cal R}(- s_{\omega})\Bigr)\Bigl(n_+(\omega)- n_-(\omega)\Bigr)\Bigr]\Bigr>~~~~\\\cr&&=
\epsilon_0\,\Lambda\,{\cal C}(\bar s)\,m_{ij},
\end{eqnarray}
where
\begin{equation}
\label{mijdef}
m_{ij}= \Bigl< d_{ij}(\omega)\,\left({n_+(\omega)- n_-(\omega)\over n_+(\omega)+ n_-(\omega)}\right)\Bigr> 
\end{equation}
is a deviatoric tensor that describes the average STZ orientation.  In analogy to Eq.(\ref{RHSsecond}), rewrite Eq.(\ref{RHSfirst}) in the form
\begin{equation}
\label{RHSfirsta}
\epsilon_0\,\Lambda\,\Bigl< d_{ij}(\omega)\,\Bigl({\cal R}(s_{\omega})-{\cal R}(- s_{\omega})\Bigr)\Bigr>
=\epsilon_0\,\Lambda\,{\cal C}(\bar s)\,{s_{ij}\over \bar s}\,{\cal T}(\bar s),
\end{equation}
where 
\begin{eqnarray}
\label{Tdef}
\nonumber
{s_{ij}\over \bar s}\,{\cal T}(\bar s) &=& \Bigl< d_{ij}(\omega)\,\left({{\cal R}(s_{\omega})-{\cal R}(- s_{\omega})\over {\cal R}(s_{\omega})+{\cal R}(- s_{\omega})} \right)\Bigr>\\\cr &\cong& {s_{ij}\over \bar s}\,
\left({{\cal R}(\bar s)-{\cal R}(- \bar s)\over {\cal R}(\bar s)+{\cal R}(-\bar s)}\right).
\end{eqnarray}

With these definitions, the plastic rate of deformation tensor in Eq.(\ref{Dpl1}) becomes
\begin{equation}
\label{Dpl2}
\tau_0\,D_{ij}^{pl}=\epsilon_0\,{\cal C}(\bar s)\,\Lambda\,\left[{s_{ij}\over \bar s}\,{\cal T}(\bar s)-m_{ij}\right].
\end{equation}
Then return to the master equation for the STZ densities, Eq.(\ref{ndot}), to deduce equations of motion for $m_{ij}$ and $\Lambda$.  Our assumption about a narrow range of relevant values of $\ell$, e.g. thresholds, means that we write $\theta_{\ell}\cong 1$, so that all the relevant STZ formation energies are of order $E_Z=k_B\,T_Z$ as anticipated in the definition of $\chi$.  Using the preceding definitions of $m_{ij}$ and $\Lambda$, we find
\begin{equation}
\label{dotm}
\tau_0\,\dot m_{ij}=2\,{\cal C}(\bar s)\,\left[{s_{ij}\over \bar s}\,{\cal T}(\bar s)-m_{ij}\right]-{m_{ij}\,\Gamma^{tot}\over\Lambda}\,e^{-1/\chi};
\end{equation}
and 
\begin{equation}
\label{dotLambda}
\tau_0\,\dot\Lambda=\Gamma^{tot}\,\Bigl(e^{-1/\chi}-\Lambda \Bigr);
\end{equation} 
where 
\begin{equation}
\label{Gamma-rho}
\Gamma^{tot} = \Gamma({\bf s}) + \rho(T).
\end{equation}

The variables $\Lambda$ and $m_{ij}$, and the roles that they play in Eqs.(\ref{Dpl2} - \ref{dotLambda}), immediately tell us a great deal about the nature of the STZ theory.  $\Lambda$ is the  fraction of molecular sites occupied by STZ's.  To be consistent with the assumption of dilute, weakly interacting STZ's, $\Lambda$ must be small, no larger than $10^{-3}$, and usually very much smaller.  The deviatoric tensor $m_{ij}$ is the average STZ orientation.  By definition, its magnitude is less than or equal to unity. In Eq.(\ref{Dpl2}), $m_{ij}$ plays the role of a back stress, consistent with the idea that STZ's already aligned in the direction of the stress impede further deformation in that direction.  The fact that the rate factor ${\cal R}(\bar s)$ is a non-negative, monotonically increasing function of its argument means that ${\cal T}(\bar s)$ is a monotonic function of $\bar s$ that vanishes when $\bar s = 0$.  Its magnitude, like that of $m_{ij}$, is bounded by unity.   
 
We also deduce from Eqs.(\ref{Dpl2} - \ref{dotLambda}) that there are two qualitatively different time scales in the STZ theory.  The plastic strain rate determined by Eq.(\ref{Dpl2}) is a rate per unit volume; it is proportional to the small quantity $\Lambda$ because it scales with the density of STZ's.  No prefactors $\Lambda$ appear in the expressions for $\dot m_{ij}$ and $\dot\Lambda$ on the right-hand sides of Eqs.(\ref{dotm}) and (\ref{dotLambda}).  These  equations describe how individual STZ's respond to changes in their environments, and the fact that  $\Lambda$ is missing as a prefactor in those equations implies that $m_{ij}$ and $\Lambda$ respond to perturbations much more rapidly than does the rate of plastic deformation.  (The factor $\exp\,(-1/\chi)/\Lambda$ on the right-hand side of Eq.(\ref{dotm}) is of order unity according to Eq.(\ref{dotLambda}).)  We have not yet written an equation of motion for the effective temperature $\chi$; but it should be clear that $\chi$ is a slow variable.  (See Sec. \ref{Teff}.) The time derivative $\dot\chi$ must be proportional to the rate per unit volume at which configurational entropy is generated during plastic deformation; thus, like $D_{ij}^{pl}$, the expression for $\dot\chi$ must contain a prefactor $\Lambda$. 

\subsection{The rate factor $\Gamma$}

Determining the rate factor $\Gamma$ that first appears here in Eq.(\ref{ndot}) has been one of the most challenging problems in the STZ theory.  This factor has been chosen incorrectly in much of the earlier literature in the field.  In our original STZ paper \cite{FL98}, Falk and I used an expression for $\Gamma$ that turns out to be correct at sufficiently low temperatures; but we did not propose a systematic rationale for it.  The problem was solved by Pechenik \cite{LP03,PECHENIK05}, who showed that $\Gamma$ can be determined by thermodynamic arguments alone.

Pechenik argued that the symmetry preserving and physically intuitive way to determine $\Gamma$ is to assume that it is proportional to the rate at which the work of deformation is dissipated irreversibly, and thus is converted into the disordered configurational fluctuations that create and annihilate STZ's.  This assumption determines $\Gamma$ uniquely if one invokes the second law of thermodynamics by requiring that this dissipation rate be non-negative.  The argument starts by writing the first law of thermodynamics in the form (using summation convention):
\begin{eqnarray}
\label{energybalance}
D_{ij}^{pl}\,s_{ij}&=&
{\epsilon_0\over\tau_0}\,{\cal C}(s)\,\Lambda\,\left[{s_{ij}\over \bar s}\,{\cal T}(\bar s)-m_{ij}\right]\,s_{ij}\cr &=& {d\over dt}\,\Psi(\Lambda,m_{ij}) + {\cal Q}(s_{ij},\Lambda,m_{ij}).
\end{eqnarray}
The left-hand side of Eq.(\ref{energybalance}) is the rate per unit volume at which plastic work is done by the stress $s_{ij}$.  On the right-hand side, $\Psi$ is the recoverable, state-dependent, internal energy density associated with the STZ's.  $\Psi$ must be proportional to the density of STZ's; therefore it is convenient to write it in the form:  
\begin{equation}
\Psi(\Lambda,m_{ij})=\epsilon_0\,\Lambda\,\psi(\bar m). 
\end{equation}
Because $\psi$ is a scalar function of just $m_{ij}$, it must depend only on $\bar m = \sqrt{(1/2)\,m_{ij}m_{ij}}$. Also assume that the effective temperature $\chi$ is so slowly varying compared to $\Lambda$ and $m_{ij}$ that it may for the moment be taken to be a constant, and need not be included as an explicit argument of the internal-energy function $\Psi$. 

The last term on the right-hand side of Eq.(\ref{energybalance}), i.e. ${\cal Q}$, is the energy dissipation rate per unit volume.  Pechenik's hypothesis is that $\Gamma$ is proportional to the rate of energy dissipation per STZ.  That is,
\begin{equation}
\label{QGamma}
{\cal Q}(s_{ij},\Lambda,m_{ij})= s_0\,{\epsilon_0 \over\tau_0}\,\Lambda\,\Gamma(s_{ij},\Lambda,m_{ij}).
\end{equation}
where $s_0$ is an as yet undetermined factor with the dimensions of stress. 

The next step is to use Eqs.(\ref{dotm}) and (\ref{dotLambda}) to evaluate the time derivatives in Eq.(\ref{energybalance}) and solve the resulting equation for $\Gamma$ or, more conveniently, $\Gamma^{tot}$.  The result is:
\begin{eqnarray}
\label{gammatotal}
&&\Gamma^{tot}(s_{ij},\Lambda, m_{ij}) = {1\over \Delta(\Lambda,\bar m)}\times \cr &&\Biggl[ {\cal C}(s)\left( {s_{ij}\over \bar s}{\cal T}(\bar s)-m_{ij} \right)\left({s_{ij}\over s_0}-{m_{ij}\over \bar m}\psi'(\bar m)\right)\cr&&+s_0\,\rho(T)\Biggr]~~~~~~~~~~~~
\end{eqnarray}
where
\begin{equation}
\Delta(\Lambda, \bar m)= s_0 -\bar m\,\psi'(\bar m)\,{e^{-1/\chi}\over\Lambda}+\psi(\bar m) \left({e^{-1/\chi}\over\Lambda}-1 \right).
\end{equation}
The quantities $\Gamma$, $\Gamma^{tot}$ and $\rho(T)$ all must be non-negative -- the first because of the second law of thermodynamics, the second and third because they are non-negative rate factors.  The condition that the numerator in Eq.(\ref{gammatotal}) remains non-negative for all values of $s_{ij}$ is sufficient to determine $\psi(\bar m)$ up to an (unnecessary) additive constant.   To find $\psi(\bar m)$, compute the inverse function of ${\cal T}$; specifically, find the function 
\begin{equation}
\label{xidef}
\xi(\bar m) = {1\over s_0}\,{\cal T}^{-1}(\bar m)
\end{equation}
such that ${\cal T}[s_0\,\xi(\bar m)]=\bar m$.  Then, because ${\cal T}$ is a monotonically increasing function of its argument, the choice $\psi'(\bar m)=\xi(\bar m)$ ensures that both $(s_{ij}/\bar s){\cal T}(\bar s)-m_{ij}$ and $(s_{ij}/s_0)-(m_{ij}/\bar m)\,\xi(\bar m)$ change sign at the same value of $s_{ij}$, and therefore that the product of these factors is never negative. 

At this stage in the development, it is useful to take advantage of the fact that $\Lambda$ relaxes so rapidly that we may replace it in the preceding formulas by its steady-state value, $\Lambda \to \exp(-1/\chi)$, and let
\begin{equation}
\Delta(\Lambda,\bar m) \to s_0\,(1 - \bar m\,\xi(\bar m)).
\end{equation}
By definition, ${\cal T}(\bar s)$ vanishes at $\bar s = 0$ and goes smoothly to $1$ as $\bar s \to \infty$.  Therefore $\bar m\,\xi(\bar m)$ vanishes like $\bar m^2$ at $\bar m=0$ and diverges at $\bar m = 1$.  So long as $\bar m$ remains in the range, $0 < \bar m < \bar m_{max}$, where $\bar m_{max}\,\xi(\bar m_{max})= 1$, the denominator $\Delta$ in Eq.(\ref{gammatotal}) remains positive as required.  Moreover, the dynamics of the system never allows an initially small $\bar m$ to reach $\bar m_{max}$ because the dissipation rate diverges at that point.  

With these simplifications, the equation of motion for $m_{ij}$, Eq.(\ref{dotm}), becomes   
\begin{eqnarray}
\label{dotm2}
&&\tau_0\,\dot m_{ij} = {1\over \Delta(\Lambda,\bar m)}\times\cr && \Bigl[{\cal C}(\bar s)\,\left({s_{ij}\over \bar s}{\cal T}(\bar s)-m_{ij}\right)\left(1-\bar m\,{\bar s\over s_0}\right)- m_{ij}\,\rho(T))\Bigr].~~~~~~~~
\end{eqnarray}
As mentioned previously, this is a stiff differential equation; there is no factor $\Lambda = \exp(-1/\chi)$ on the right-hand side to produce slow relaxation.  Therefore, for most purposes, we can set $\dot m_{ij} = 0$ and replace $m_{ij}$ elsewhere by its stress dependent value for which the right-hand side of Eq.(\ref{dotm2}) vanishes.  Moreover, for an isotropic system in which the only orientations are set by the stress tensor $s_{ij}$, we must have 
\begin{equation}
\label{mijM}
m_{ij} \to {s_{ij}\over \bar s}\,M(\bar s)
\end{equation}
where $M(\bar s)$ is the stationary solution of Eq.(\ref{dotm2}) with the preceding ansatz.  Specifically,
\begin{eqnarray}
\label{M-s}
\nonumber
&&M(\bar s)={s_0\over 2\,\bar s}\,\left[1+ {\bar s\over s_0}\,{\cal T}(\bar s)+{\rho(T)\over 2\,{\cal C}(\bar s)}\right]\cr\\ &&- {s_0\over 2\,\bar s}\sqrt{\left[(1+ {\bar s\over s_0}\,{\cal T}(\bar s)+{\rho(T)\over 2\,{\cal C}(\bar s)}\right]^2 - 4\,{\bar s\over s_0}\,{\cal T}(\bar s)}\,.~~~~~~~~~
\end{eqnarray}
We then have
\begin{equation}
\label{Dpl3}
\tau_0\,D_{ij}^{pl}(\bar s,\chi)=\epsilon_0\,{\cal C}(s)\,e^{-1/\chi}\,{s_{ij}\over \bar s}\,\bigl[{\cal T}(\bar s)-M(\bar s)\bigr],
\end{equation}
which has the form anticipated in Eq.(\ref{D-chi-s}).

For later reference:
\begin{equation}
\label{gammatilde3}
\Gamma^{tot}(\bar s)={2\,{\cal C}\,\bigl[{\cal T}-M\bigr]\,\bigl[(\bar s/ s_0)-\xi(M)\bigr]+\rho(T)\over 1-M\,\xi(M)};
\end{equation}
and
\begin{equation}
\label{Gamma}
\Gamma(\bar s)={2\,{\cal C}\,\bigl[{\cal T}-M\bigr]\,\bigl[(\bar s/ s_0)-\xi(M)\bigr)+M\,\xi(M)\,\rho(T)\over 1-M\,\xi(M)},
\end{equation}
where ${\cal C}$, ${\cal T}$, and $M$ are all understood to be functions only of $\bar s$ and $T$.  The last expression reduces to the formula for $\Gamma(\bar s)$ postulated in \cite{FL98}, but it does so only for temperatures low enough that $\rho(T)=0$ (i.e. below the glass transition), and that reverse STZ transitions are negligible, ${\cal R}(-\bar s) \approx 0$.  Then ${\cal T}(\bar s) \approx 1$ and $\xi(M) \approx 0$, so that 
\begin{equation}
s_0\,{\epsilon_0\over \tau_0}\,e^{-1/\chi}\,\Gamma(\bar s) \approx D_{ij}^{pl}s_{ij}.
\end{equation}
This relation between the STZ production rate and the applied power density has been confirmed recently by Heggen et al. \cite{HEGGENetal05} in the context of conventional flow-defect theories.

\section{The dynamic yield stress}
\label{yieldstress}

One of the most notable features of the STZ theory is the natural way in which a yield stress emerges from its basic structure.  To see this, consider the case where the bath temperature $T$ is below the glass transition, so that the spontaneous STZ annihilation and creation rates, proportional to $\rho(T)$, both vanish.  This condition does not necessarily mean that the transition rate ${\cal R}(-\bar s)$ vanishes, as assumed in the strictly athermal version of the STZ theory.\cite{BLP07I,BLP07II} Thermal fluctuations may induce backward STZ transitions even when the system is in a completely inviscid glassy state, i.e. when it has infinite linear shear viscosity -- which is the only situation in which there is a sharply defined yield stress.  

Setting $\rho(T)=0$ in Eq.(\ref{dotm2}), or simply evaluating Eq.(\ref{M-s}) in this limit, we find $M(\bar s)$ to be
\begin{equation}
\label{mtilde2}
M(\bar s)\to \cases{{\cal T}(\bar s)& for $\bar s < s_y$, \cr s_0/\bar s & for $\bar s > s_y$,}
\end{equation}
where  $s_y$ is the value of $\bar s$ at which the two branches of $M(\bar s)$ cross; that is, $s_y$ is the solution of 
\begin{equation}
\label{sy}
s_y\,{\cal T}(s_y)=s_0.
\end{equation}
Clearly, $s_y$ is a dynamic yield stress.  When $\bar s<s_y$, Eqs.(\ref{Dpl3}) and (\ref{mtilde2}) imply that $D_{ij}^{pl}=0$; the system is jammed and no plastic deformation is occurring.  When $\bar s > s_y$, $D_{ij}^{pl}$  is nonzero.  The two cases of the function $M(\bar s)$ shown in Eq.(\ref{mtilde2}) are the stable branches of the steady-state solutions of Eq.(\ref{dotm2}); thus the yield stress occurs at an exchange of stability.  Note that the denominator on the right-hand side of Eq.(\ref{dotm2}) vanishes only at $\bar m= s_0\,/s_y =  {\cal T}(s_y)$, that is, at the value of $\bar m$ corresponding to the exchange of stability at $\bar s = s_y$; therefore $\bar m$ is dynamically constrained to remain between these limits as anticipated in the discussion preceding Eq.(\ref{dotm2}).  

As defined here, the yield stress $s_y$ is an intrinsic, steady-state  property of a material; it does not depend on the history of deformation or even on the material's initial state of disorder.  This quantity  can appropriately be called the ``ultimate yield stress'' or the ``minimum flow stress.''  It is quite different from the ``peak stress,'' which will appear in the transient stress-strain curves computed in Sec.\ref{metglass}, and which has no such intrinsic meaning.

Equation (\ref{sy}) unambiguously determines the proportionality factor $s_0$ that appears in the Pechenik relation, Eq.(\ref{QGamma}), in terms of the yield stress $s_y$.  Because $s_y$ generally depends on temperature, pressure, and perhaps other state variables, $s_0$ must also depend on those variables. The temperature dependence of $s_0$ may be especially important near a glass transition, where thermal fluctuations become increasingly effective in assisting molecular motions over energy thresholds, so that $s_y$ is a decreasing function of increasing $T$.  In fact, $s_0$ is a more fundamental quantity than $s_y$.  It remains well defined above the glass transition, where Eq.(\ref{sy}) is no longer valid and where, strictly speaking, there is no yield stress, but a function such as $D_{ij}^{pl}(\bar s,\chi)$ in Eq.(\ref{Dpl3}) may still exhibit visible stress dependence as $\bar s$ increases through $s_0$. 

In the literature on flow-defect theories of amorphous plasticity, it is generally assumed that -- in the language of the STZ theory -- the yield stress is determined by some intrinsic feature of the transition rate ${\cal R}(\bar s)$.  Supposedly, ${\cal R}(\bar s)$ rises sharply when the stress grows large enough to drive molecules in an STZ over an energy barrier, and the barrier stress is the yield stress $s_y$.  This assumption cannot be correct.  For example, in the low-temperature limit mentioned at the end of Sec.\ref{STZequations}, where ${\cal R}(-\bar s)\approx 0$ , Eq.(\ref{Tdef}) tells us that ${\cal T}(\bar s) \approx 1$, so that $s_y \approx s_0$, independent of whatever function we choose for ${\cal R}(\bar s)$ and without any necessary relation to the dynamic ingredients of the rate factor.  

How, then, are we to understand $s_0$?  The easy answer is that we obtain it from experiment or calculate it in some approximate way, and then build it into our choice of ${\cal R}(\bar s)$.  The chosen ${\cal R}(\bar s)$ determines the behavior of $D_{ij}^{pl}(\bar s,\chi)$, and thus fixes the relation between stress and plastic flow at all stresses, including near $s_0$.  But, had we  chosen a ``yield stress'' different from $s_0$ in our choice of ${\cal R}(\bar s)$, we would not have changed Eq.(\ref{sy}) very much or obtained a significantly different value of $s_y$.  In fact, we do not know whether any characteristic stress appearing in ${\cal R}(\bar s)$ is related to $s_0$.  Therefore, simply inserting a physically motivated choice of $s_0$ into both ${\cal R}(\bar s)$ and Eq.(\ref{sy}), while perhaps sensible, evades the basic question.  In principle, we need independent estimates of both $s_0$ and ${\cal R}(\bar s)$.

To find an independent estimate of $s_0$, return to the original Pechenik relation, Eq.(\ref{QGamma}).  The left-hand side of this equation is ${\cal Q}= T\,dS/dt$, where $dS/dt$ is the rate per unit volume at which the entropy $S$ is increasing as a result of plastic deformation.  On the right-hand side, writing $\Lambda = \exp(-1/\chi)$, identify
\begin{equation}
\label{Scprod}
{\epsilon_0\over\tau_0}\,e^{-1/\chi}\,\Gamma(\bar s)\cong {\Omega\over k_B\,\nu_Z}\,\left({dS_c\over dt}\right)_{mech}
\end{equation}
where $(dS_c/dt)_{mech}$ is the rate at which configurational entropy in the form of STZ's is being produced during mechanical deformation, and $\Omega \sim n_{\infty}^{-1}$ is the volume per molecule.  The factor $k_B\,\nu_Z$ is an approximation for the entropy of an STZ, which means that $\nu_Z$ is roughly the number of molecules in an STZ.  Thus Eq.(\ref{QGamma}) takes the form
\begin{equation}
\label{s0-S}
{\cal Q} = T\,{dS\over dt}= {s_0\,\Omega\over k_B\,\nu_Z}\,\left({dS_c\over dt}\right)_{mech}.
\end{equation}
From this we deduce that the fraction of the total dissipation rate ${\cal Q}$ that produces STZ-like configurational disorder is  $\nu_Z\,k_B\,T/s_0\,\Omega$.   In other words, an amorphous material yields more easily if it converts a larger fraction of the work of deformation into configurational disorder instead of ordinary heat.  

The relation between $s_0$ and entropy generation in Eq.(\ref{s0-S}) is not purely formal; it has physical content, and therefore should be useful for evaluating $s_0$.  To illustrate this possibility, we can make rough estimates of both sides of Eq.(\ref{s0-S}).  For this purpose, assume that the temperature is very small, well below the glass transition.  On  the left-hand side of Eq.(\ref{s0-S}), consider any irreversible molecular rearrangement driven by an applied shear stress.  The system first deforms elastically as the molecules are driven to an unstable threshold, and then all of this elastic energy is converted into some combination of kinetic energy (heat) and configurational energy as the system relaxes to its rearranged state of equilibrium.  The local shear strain at threshold must be of order unity; thus we have $T\,dS/dt \sim \mu\,\dot\gamma$, where $\mu$ is the shear modulus, and the total shear rate $\dot\gamma$ is a measure of the rate at which these rearrangements are occurring.  To evaluate the right-hand side of Eq.(\ref{s0-S}), note that this term is equal to the rate at which STZ's are being created, each adding an entropy increment of order $k_B$ per molecule. Again, assume that the relevant low-temperature rate of events is proportional to $\dot\gamma$, so that $\Omega\,(dS_c/dt)_{mech} \sim k_B\,\nu_Z\,\dot\gamma$.  

With these estimates, Eq.(\ref{s0-S}) becomes simply
\begin{equation}
\label{s0-mu}
s_0 \sim {\mu\over \nu_Z}.
\end{equation}
This is the relationship found by Johnson and Samwer \cite{JOHNSON-SAMWER05}, who pointed out that the low-temperature yield strain $s_y/\mu$ is an almost universal constant, of order $0.03$, for thirty different bulk metallic glasses.  A literal interpretation of  Eq.(\ref{s0-mu}) would imply that $\nu_Z \sim 30$, which may be reasonable for these complex, multicomponent materials.  However, Eq.(\ref{s0-mu}) is not much more than a dimensional analysis, and should not be taken so seriously.  A complete calculation would include a theory of the relation between ${\cal Q}$ and the elastic driving forces, and would require a better estimate of the relation between the size of an STZ and its entropy.  Both of those calculations are well beyond the scope of this paper. 

\section{Newtonian linear viscosity}
\label{viscosity}

It is convenient at this point to use the STZ equations derived so far to obtain an expression for the Newtonian linear viscosity.  Consider the case of temperatures just above the glass transition, where $\rho(T)$ is small but nonzero, and compute the plastic strain rate in the limit of vanishingly small driving stress $s$.  The quantity $M(\bar s)$ defined in Eq.(\ref{M-s}) must vanish linearly as $\bar s \to 0$, so the term $\bar m\,\bar s/s_0$ in the numerator of Eq.(\ref{dotm2}) may be neglected and, with Eq.(\ref{mijM}), the equation for $M(\bar s)$ becomes
\begin{equation}
\label{dotmapprox}
2\,{\cal C}(\bar s)\,\Bigl({\cal T}(\bar s)-M\Bigr) \cong M\,\rho(T).  
\end{equation}
Therefore,
\begin{equation}
\label{mtildeapprox}
M(\bar s)\cong {{\cal T}(\bar s)\over 1 + {\rho(T)\over 2\,{\cal C}(\bar s)}} \approx {{\cal T}'(0)\,\bar s \over 1+ {\rho(T)\over 2\,{\cal C}(0)}}.
\end{equation}
For steady-state motion at temperatures above the glass transition, in the limit of infinitesimally small deformation rate, the effective temperature $T_{e\!f\!f}$ becomes the bath temperature $T$, so that $\chi \to T/T_Z$.  Using Eqs.(\ref{Dpl3}), (\ref{dotmapprox}), and  (\ref{mtildeapprox}), we find (for pure shear) 
\begin{equation}
D_{xx}^{pl}= - D_{yy}^{pl}\approx {\epsilon_0\over 2\,\tau_0}\,{{\cal T}'(0)\,s \over 1+ {\rho(T)\over 2\,{\cal C}(0)}}\,\rho(T)\,e^{-T_Z/T}.
\end{equation}

The Newtonian viscosity is 
\begin{eqnarray}
\label{etaN}
\nonumber
&&\eta_N(T) \equiv \lim_{\bar s \to 0}{\bar s\over 2 D_{xx}^{pl}}= {\tau_0\over \epsilon_0\,{\cal T}'(0)}\,\left[1+ {\rho(T)\over 2\,{\cal C}(0)}\right]\,{e^{T_Z/T}\over \rho(T)}\cr\\ &&\approx {\tau_0\over \epsilon_0\,{\cal T}'(0)}\,{e^{T_Z/T}\over \rho(T)}= \eta_0\,\exp\,\left[{T_Z\over T}+\alpha(T)\right],
\end{eqnarray}
where $\eta_0^{-1} = (\epsilon_0/\tau_0)\,{\cal T}'(0)$, and the super-Arrhenius function $\alpha(T)$ is defined in Eq.(\ref{alphadef}). The approximation in Eq.(\ref{etaN}) is valid only at temperatures close enough to the glass transition that $\rho(T)\ll {\cal C}(0)$, and thus neglects a potentially important  temperature dependence of the prefactor $\eta_0$ at larger $T$.  

Note that, especially in its limiting form near the glass transition, the viscosity is determined almost entirely by the super-Arrhenius rate of thermally activated rearrangements and not by the STZ transition rate itself.  In that connection, note also that I am making a different assumption here than the one I used in \cite{JSL04}, where I guessed that the rate factor ${\cal R}(\bar s)$ vanished super-Arrheniusly at the glass transition, and that $\rho(T)/ {\cal C}(0)\to 1$ at that point.  That assumption was inconsistent with the fact that, even though their linear viscosities vanish, glassy materials do undergo plastic deformation at low temperatures and large driving forces.  (See the discussion at the beginning of Section \ref{metglass}.)  

\section{Effective-temperature dynamics}
\label{Teff}

We now need an equation of motion for the dimensionless effective temperature $\chi$.  There is an extensive literature on this subject.  References that I have found useful include \cite{CUGLIANDOLOetal97,SOLLICHetal97,BERTHIER-BARRAT,OHERNetal04,ILG-BARRAT07}.  Much of what follows is based directly on work by Liu and colleagues, especially \cite{ONOetal02,HAXTON-LIU07}. 

As in \cite{JSL04}, the equation of motion for $\chi$ to be used here is an approximate statement of heat balance. It has the form
\begin{eqnarray}
\label{dotchi1}
\nonumber
C_{e\!f\!f}\,T_Z\,\dot \chi &=& T_{e\!f\!f}\,\left({dS_c\over dt}\right)_{mech}\,\left[1 - {\chi\over \hat\chi(q)}\right]\cr\\ & + & T_{e\!f\!f}\,\left({dS_c\over dt}\right)_{therm}\,\left[1 - {\chi\,T_Z\over T}\right].~~~~
\end{eqnarray}

On the left-hand side, $C_{e\!f\!f}$ is an effective specific heat per unit volume of the form $C_{e\!f\!f} = k_B\, c_0/\Omega$, where $c_0$ is a dimensionless number of order unity.  Thus this term is a rough estimate of the rate at which the configurational heat content is changing as a function of time. 

The two terms on the right-hand side of Eq.(\ref{dotchi1}) are the rates at which the configurational heat content is being changed, respectively, by mechanical work and by thermal fluctuations.  The factor $(d\,S_c/dt)_{mech}$, introduced in Eq.(\ref{Scprod}), is the rate at which configurational entropy is produced in the form of STZ's during mechanical deformation.  The definition of the effective temperature implies that we multiply this rate by $T_{e\!f\!f}$ to obtain the heat production.  In the second term, $(d\,S_c/dt)_{therm}$ denotes the rate of configurational entropy production by thermally induced molecular rearrangements.  This term is closely related to the thermal STZ creation rate, proportional to the function $\rho(T)$ introduced in Eq.(\ref{ndot}); but it requires further consideration.  

Both the factors $(1-\chi/\hat\chi)$ and $(1- \chi\,T_Z/T)$ appearing in Eq.(\ref{dotchi1}) introduce physical mechanisms that are beyond the basic STZ assumptions.  The second is the easiest to understand.  It says simply that, at temperatures above the glass transition, the effective temperature  $\chi$ relaxes toward $T/T_Z$; and it models this aging effect by a conventional linear law of cooling.  The first factor,  i.e. the modification of the mechanically driven rate of entropy production, is the more interesting.  It says that, at low temperatures where thermal fluctuations are negligible, $\chi$ relaxes to a steady-state value denoted here by $\hat\chi(q)$, and it assumes that this relaxation -- like the cooling law used in the thermal term -- is linear.

To interpret the mechanical factor, think about bath temperatures below the glass transition, so that molecular rearrangements are not spontaneously activated by thermal fluctuations but must be driven by externally applied forces.  Those forces, in effect, ``stir'' the system at a rate, say, $\dot\gamma$, which may be the norm of the rate-of-deformation tensor. The only intrinsic time scale in the system at low temperatures is $\tau_0$; therefore, define the dimensionless stirring rate to be $q\equiv\dot\gamma\,\tau_0$.  Consider first the limit $q \ll 1$.  If, in order to achieve steady-state statistical equilibrium, each molecule must have changed its neighbors at least once or twice, then the equilibration time is irrelevant; only the magnitude of the deformation makes a difference.  After long, slow stirring, the state of the system is characterized by an effective disorder temperature that is independent of the stirring rate or any details of the stirring mechanism.  Denote this value of $\chi$ by $\chi_0$; that is, $\hat\chi \to \chi_0$ as $q \to 0$. (In earlier papers, e.g. \cite{JSL04,BLP07II,MLC07}, $\chi_0$ was denoted by $\chi_{\infty}$.)

More generally, if $q$ is not negligibly small, then the steady-state value of $\chi$ is  $\hat\chi(q)$.  The $q$-dependence of $\hat\chi$ is likely to be nontrivial for foams or granular materials, where $\tau_0$ is not microscopically small, and experimentally accessible strain rates may be comparable to $\tau_0^{-1}$.  This situation also may arise in amorphous molecular materials when strain rates are very large, for example, near crack tips or at the centers of shear bands.  As in \cite{JSL-MANNING07}), the $q$ dependence of $\hat\chi$ may be quite interesting.

The linear approximations made in both of the terms on the right-hand side of Eq.(\ref{dotchi1}) imply that this equation cannot be  used to describe very large excursions from steady-state equilibria. $\chi$ cannot be far from $\hat\chi(q)$ when $(d\,S_c/dt)_{mech}$ is large and $(d\,S_c/dt)_{therm}$ is  small; nor can $\chi$ be very far from $T/T_Z$ in the opposite situation.  

Return now to the entropy-production terms in Eq.(\ref{dotchi1}).  The factor $(d\,S_c/dt)_{mech}$ as given in Eq.(\ref{Scprod}) appears here, because the mechanical generation of configurational entropy requires the STZ mechanism. The problem of evaluating the thermal factor $(d\,S_c/dt)_{therm}$ is more problematic and interesting.  By definition, the effective temperature pertains to all the configurational degrees of freedom of the system, not just those that couple to external shear stresses.  Above the glass transition, thermal fluctuations may generate a variety of different kinds of defects, with formation energies different from $k_B\,T_Z$, and these will contribute to the configurational entropy.  With this possibility in mind, but keeping as close an anology to Eq.(\ref{Scprod}) as possible, I propose that the thermal factor have the form:
\begin{equation}
\label{Sctherm}
T_{e\!f\!f}\,\left({d\,S_c\over dt}\right)_{therm}= \kappa\,{\nu_Z\,\epsilon_0\,k_B\,T_{e\!f\!f}\over \tau_0\,\Omega}\,\rho(T)\,e^{-\beta/\chi}.
\end{equation}
Here, I have replaced the mechanical factor $\Gamma(\bar s)$ in Eq.(\ref{Scprod}) by the thermal factor $\rho(T)$ on the assumption that these functions play the same roles as noise strengths as they did in Eq.(\ref{ndot}).  To account for the possibility that non-STZ configurational disorder may be included in the entropy production, I have replaced $\exp\,(-1/\chi)$ by $\kappa\,\exp\,(-\beta/\chi)$, where $\beta\,k_B\,T_Z$ is an activation free energy for defect formation, and $\kappa$ is a dimensionless rescaling of the factor $\nu_Z/\Omega$.  Note that the parameters $\beta$ and $\kappa$ play the same roles here as they did in \cite{JSL04}; but now it is more obvious why  $\beta$ should be smaller than unity, because it is a rough approximation for a free energy that may include more disordered  states than those described just by STZ's.  In  \cite{JSL04}, I showed that systems with $\beta < 1$ may be especially susceptible to shear-band formation. 

With the preceding assumptions, Eq.(\ref{dotchi1}) becomes
\begin{eqnarray}
\label{dotchi2}
\nonumber
\tau_0\,\tilde c_0\,\dot\chi &=& e^{-1/\chi}\,\Gamma(\bar s)\,\chi\,\left[1-{\chi\over \hat\chi(q)}\right]\cr\\&+&\kappa\,e^{-\beta/\chi}\,\rho(T)\,\chi\,\left[1-{T_Z\,\chi\over T}\right],~~~~~
\end{eqnarray}
where $\tilde c_0 = (c_0/\epsilon_0\,\nu_Z)$.  $\Gamma(\bar s)$ is shown in Eq.(\ref{Gamma}).  This form of the $\dot\chi$ equation differs slightly from its original version, Eq.(3.5) in \cite{JSL04}.  Importantly, use of Eq.(\ref{Scprod}) simplifies the final result and eliminates the need for an extra approximation in the derivation. Note that $\hat\chi(q)$ naturally appears in the denominator inside the first bracketed term on the right-hand side of Eq.(\ref{dotchi2}). Thus, large values of $\hat\chi$ do not produce unphysical behavior as would have happened previously.  

The effective temperature is emerging as a remarkably powerful concept for understanding the nonequilibrium properties of amorphous materials.  It plays a key role in the analysis of large-scale deformation of metallic glasses, first published in \cite{JSL04}, and reworked here in Section \ref{metglass}.  In that case, the dynamics of $\chi$ as determined by Eq.(\ref{dotchi2}) control the plastic response of the system to transient changes in external driving.  I know of no other mechanism capable of quantitatively explaining the observed relaxation phenomena.

Some of the most interesting recent developments indicate that, in steady-state nonequilibrium situations, $\chi$ can be used ``quasithermodynamically'' as if it were an ordinary temperature. That is, $\chi$ can be used as an independent intensive variable, along with the thermal temperature, the pressure, etc., in equations of state for extensive quantities such as the volume or internal energy.  

Perhaps the first example of a quasithermodynamic analysis is that of Shi {\it et al.} \cite{SHIetal07}, who postulated a linear equation of state for the potential energy as a function of $\chi$ and used this to interpret their molecular-dynamics simulations of shear banding.  Manning, Carlson, and I \cite{MLC07} used Eq.(\ref{dotchi2}) and the data in \cite{SHIetal07} to 
show that shear banding occurs here via a nonlinear, transient instability.  Another quasithermodynamic analysis appears in a paper by Bouchbinder, Procaccia and myself \cite{BLP07II}, in which we used an athermal version of STZ theory to interpret molecular dynamics simulations of amorphous silicon by Demkowicz and Argon.\cite{DA04,DA05-1,DA05-2,AD06}  The latter authors measured not just stress-strain curves but also the fraction of atoms that were in liquidlike, as opposed to solidlike, nearest-neighbor configurations.  We found that this liquidlike fraction, in steady-state nonequilibrium situations, obeys a quasithermodynamic equation of state as a function of $\chi$. With this understanding, we were able to account quantitatively for the simulation results, including the time dependent transients in the liquidlike fraction that Demkowicz and Argon observed near the onset of loading.

The most strikingly unexpected and speculative quasithermodynamic role played by $\chi$ is suggested by the molecular dynamics simulations of Haxton and Liu.\cite{HAXTON-LIU07}  These authors simulated a simply sheared, two-dimensional, glass-forming material over three decades of steady-state strain rates $\dot\gamma$ and for bath temperatures $T$ ranging from about one tenth of the glass transition temperature $T_0$ to about twice $T_0$.  By measuring pressure fluctuations, they independently determined values of $\chi$ for each value of $\dot\gamma$ and $T$.  For $T < T_0$, Manning and I \cite{JSL-MANNING07} propose that their observed relation between $\dot\gamma$ and $\chi$ is a direct analog of the  relation between the $\alpha$ relaxation rate and the bath temperature $T$ near a conventional glass transition.  More specifically, the relation between the dimensionless strain rate $q$ and the low-temperature ($\rho = 0$), steady-state $\chi = \hat\chi(q)$ in Eq.(\ref{dotchi2}) can be written in the form 
\begin{equation}
\label{q-chi}
\frac{1}{q(\hat\chi)} = {1 \over q_0} \exp \left[ {A\over \hat\chi} + \alpha_{e\!f\!f}(\hat\chi)\right],
\end{equation}
which is a direct analogy to the final form of the Newtonian viscosity in Eq.(\ref{etaN}).  Here, $q_0^{-1}$ and $A$ are constants analogous to $\eta_0$ and $T_Z$.  $\alpha_{e\!f\!f}(\hat\chi)$ is a super-Arrhenius function that has the same form as $\alpha(T)$ in Eqs.(\ref{alphadef}) and (\ref{etaN}), diverging at an effective Kauzmann temperature $\chi_0$ and vanishing above some $\chi_A$.  In other words, Manning and I interpret $q$ to be a dimensionless rate at which molecular rearrangements are driven by $\chi$ fluctuations below $T_0$.  According to this extreme version of the quasithermodynamic hypothesis, $q$ should exhibit the same behavior -- including a glass transition at $\chi_0$ -- as that which occurs for thermally driven rearrangements above $T_0$. Another interesting prediction of this hypothesis is that $\hat\chi$ diverges at a finite value of the dimensionless strain rate, i.e. at $q=q_0$.  If true, the material would ``melt,'' and the solidlike STZ assumptions would fail at a well defined upper limit of the driving strength.  
  
\section{ STZ theory of deformation in a bulk metallic glass}
\label{metglass}

To date, the most complete applications of the STZ theory in the interpretation of laboratory experiments have been our analyses \cite{FLP04,JSL04} of the metallic glass data published by Lu et al.\cite{LUetal03}.  These papers were written before we understood several crucial aspects of the theory.  For example, we had not recognized that the separation of time scales discussed in Section \ref{STZequations} implies that the parameter $\epsilon_0$ never appears by itself in the equations of motion, but always in the combination $\epsilon_0/\tau_0$.  As a result, the values of $\epsilon_0$ and $\tau_0$ cited in those papers were dramatically incorrect, although the errors mostly cancelled out in the final results.  In the present interpretation, $\epsilon_0 \sim 1$, and $\tau_0$ is of order femtoseconds; thus $\epsilon_0/\tau_0 \sim 10^{15}\,{\rm sec}^{-1}$.  

Another example is that, in those earlier papers, we had not understood the relationship -- or lack thereof -- between the STZ transition rate ${\cal R}(s)$ and the yield stress, and therefore were not thinking carefully enough about the physical basis of ${\cal R}(\bar s)$.  We can now write an expression for this rate factor that is directly related to molecular mechanisms, without being constrained to include the yield stress explicitly in that formula.  Moreover, the excitation-chain theory of anomalously slow relaxation near the glass transition \cite{JSL-XCPRE06,JSL-XCPRL06} implies that the super-Arrhenius factor $\rho(T)$  determines the rates of spontaneous creation and annihilation of STZ's and thermalization of the effective temperature, but does not belong in ${\cal R}(\bar s)$ itself -- as was assumed incorrectly in \cite{JSL04}.  

It is useful, therefore, to conclude this paper by revisiting the metallic glass data, both to update the earlier papers and to show how the theoretical ideas developed in this one can be brought to bear on real-world phenomena.

Lu et al.\cite{LUetal03} measured deformations of the bulk metallic glass Vitreloy 1, ${\rm Zr_{41.2}\,Ti_{13.8}\,Cu_{12.5}\,Ni_{10}\,Be_{22.5}}$.  To a good approximation, their system consisted of a uniform bar with uniaxial compressive stress applied, say, in the $x$ direction, and stress-free surfaces normal to the $y$ and $z$ directions.  This total stress tensor $\sigma_{ij}$ has only one nonzero element, $\sigma_{xx} \equiv \sigma$, which is the experimentally reported stress.  Therefore $\sigma = (3/2)\,s_{xx} = \sqrt{3}\,\bar s$.  Similarly, the measured plastic strain rate is $D_{xx}^{pl} = (2/\sqrt{3})\,{\bar D}^{pl}$, where ${\bar D}^{pl}$ is the magnitude of the plastic rate-of-deformation tensor $D_{ij}^{pl}$ in Eq.(\ref{Dpl3}).  The Newtonian viscosity is 
\begin{equation}
\label{etaN2}
\eta_N \equiv \lim_{s_{xx} \to 0}{s_{xx}\over 2 D_{xx}^{pl}}=\lim_{\bar s \to 0}{\bar s\over 2 \bar D^{pl}},
\end{equation}
therefore Eq.(\ref{etaN}) remains unchanged.  

We next must specify the STZ transition rate ${\cal R}(\bar s)$.  For the moment, return to a notation in which $\bar s = s$ and $\bar m = m$ have signs, because backward transitions with $s < 0$ play a role in the analysis, as is clear in Eqs.(\ref{Tdef}).  In earlier versions of the STZ theory, we always chose the simplest possible forms of this transition rate on the assumption that whatever data we had available from experiments or simulations would not justify additional theoretical complications.  Manning and I departed from this purely phenomenological approach in \cite{JSL-MANNING07}, primarily because the Haxton-Liu data \cite{HAXTON-LIU07} that we were interpreting extended over an exceptionally wide range of stresses and strain rates, and therefore required us to construct and test a model that was more physically motivated than the earlier ones.  

The transition rate that Manning and I used in \cite{JSL-MANNING07}, which I will adopt here, includes an Eyring-like activation factor at small stresses, and a smooth transition from Eyring to power-law behavior at large stresses.  Specifically,
\begin{equation}
\label{R-s}
R(s) = \exp \left[-{T_E\over T}\,e^{-s/\tilde\mu}\right]\left[1 + \left({s\over s_1}\right)^{2} \right]^{n/2},
\end{equation}
We then have:
\begin{eqnarray}
{\cal C}(s) &=& \exp\left[-{T_E\over T}\,\cosh (s/\tilde\mu)\right]\,\times\cr && \cosh\left[{T_E\over T}\,\sinh (s/\tilde\mu)\right]\,\left[1 + \left({s\over s_1}\right)^{2} \right]^{n/2},
\end{eqnarray}
and
\begin{equation}
{\cal T}(s)= \tanh\left[{T_E\over T}\,\sinh (s/\tilde\mu)\right],
\end{equation}
so that
\begin{equation}
\xi(m)= \tilde\mu\,{\rm arcsinh}\left[{T\over T_E}\,{\rm arctanh}(m)\right].
\end{equation}

The first factor on the right-hand side of Eq.(\ref{R-s}) is the Eyring rate in a form similar to that used in \cite{FL98}, where the exponential function of $s/\tilde\mu$ causes the rate to saturate at large $s$.  The energy $k_B\,T_E$ is the height of the Eyring activation barrier; and we expect $T_E/T_Z < 1$.  The parameter $\tilde\mu$ is the stiffness of the barrier at its peak.  It appears here ostensibly in the same place that $\bar\mu$ appeared in \cite{JSL04}, but now it has no direct connection to the yield stress. The Eyring factor explicitly expresses the fact that,  even at temperatures below the glass transition, the barrier opposing STZ transitions must be appreciably smaller than the STZ formation energy.  In the limit of small stress, $s\ll \tilde\mu$:   
\begin{equation}
{\cal R}(s) \approx \exp \left[-{T_E\over T}\,(1 - s/\tilde\mu)\right];
\end{equation}
therefore
\begin{equation}
\label{Csmall}
{\cal C}(s) \approx e^{-T_E/T} \cosh\left({T_E\,s\over T\,\tilde\mu}\right);
\end{equation}
and
\begin{equation}
\label{Tsmall}
{\cal T}(s) \approx \tanh \left({T_E\,s\over T\,\tilde\mu}\right).
\end{equation}

The second factor on the right-hand side of Eq.(\ref{R-s}) converts the saturated Eyring function at large $s$ to a power law:
\begin{equation}
\label{Csn}
{\cal C}(s) \approx \left({s\over s_1}\right)^n;~~~~{\cal T}(s) \approx 1.
\end{equation}
Here, $s_1$ is a stress scale that should be of order $s_y$; that is, the crossover between Eyring behavior and large-scale plasticity should occur near the yield stress. The power law has a physical interpretation.  In dissipative systems, and when strain rates are not too large ($\dot\gamma\,\tau_0 \ll 1$), we expect $n = 1$, corresponding to linear friction or viscosity.  In the opposite situation, as in \cite{HAXTON-LIU07,JSL-MANNING07}, where rates are large, and the motion is controlled by hard-core collisions rather than the details of molecular interactions, we find Bagnold scaling \cite{BAGNOLD} with $n = 1/2$.  That is, if there is no natural energy scale in the problem, then dimensional analysis requires that the stress be proportional to the square of a rate; and if ${\cal R}(s)/\tau_0$ is the only available quantity with dimensions of inverse time, we have ${\cal R}(s) \sim s^{1/2}$.  In the experiments of Lu et al, stresses never exceed the yield stress, and strain rates are always very much less than $\tau_0^{-1}$; therefore, from here on, I use the linear law with $n=1$.

\begin{figure}[h]
\centering \includegraphics[height=8 cm]{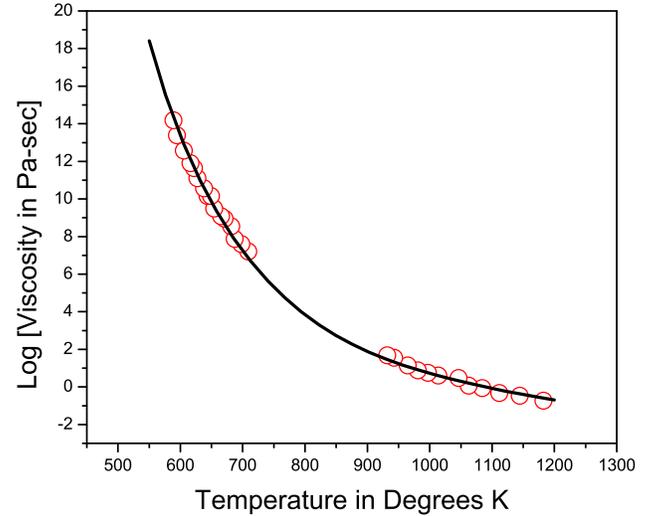}
\caption{\label{etaMG} (Color online) Viscosity of Vitreloy 1. The red circles are the experimental measurements reported in \cite{LUetal03}.  The solid curve is the theoretical fit described in the text following Eq.(\ref{etaN3}). }
\end{figure}

With this choice of the STZ transition rate, the Newtonian viscosity in Eq.(\ref{etaN}) becomes
\begin{equation}
\label{etaN3}
\eta_N(T) = {\tau_0\,\tilde\mu\,T\over \epsilon_0\,T_E}\,\left[1 + {1\over 2}\,e^{T_E/T-\alpha(T)}\right]\,e^{T_Z/T + \alpha(T)}.
\end{equation}
For simplicity,  I use the approximation
\begin{equation}
\label{alphaTB}
\alpha(T)\cong {T_1\over T-T_0}\,e^{- a\,{T-T_0\over T_A-T_0}},
\end{equation}
which makes a smooth transition from Vogel-Fulcher behavior near $T_0$ to Arrhenius behavior above $T_A$. Fig.\ref{etaMG} contains the viscosity data for Vitreloy 1 shown in Fig.10 of \cite{LUetal03}.  As pointed out by Masuhr et al \cite{MASUHRetal99}, this data does not fit neatly into the usual Vogel-Fulcher scheme. In fact, it is more easily fit by the Cohen-Grest formula \cite{COHEN-GREST79}, which contains no low temperature divergence, and which is what I used in \cite{JSL04}.  However, the data can be fit by Eq.(\ref{etaN3}), which is consistent with other parts of the present analysis.  

The solid curve in Fig.\ref{etaMG} has been plotted using $\tau_0\,\tilde\mu/\epsilon_0 = 10^{-7}$ Pa-sec., $T_0 = 250\,K$, $T_Z= 16,000\,K$, $T_E = 3,000\,K$, $T_A = 1,000\,K$, $T_1 = 31,000\,K$ and $a = 3$.  Note that the glass transition temperature $T_0$ is very small, consistent with the Cohen-Grest analysis, and that $T_1$ needs to be remarkably large in order to fit the data.  (Note also, however, that the effective value of $T_1$ in the range of interest is very much smaller because of the exponential cutoff.)  The ratio $T_E/T_Z \sim 0.2$ seems reasonable, i.e. the Eyring barrier is substantially smaller than the STZ formation energy but has roughly the same energy scale.  Similarly, if $\epsilon_0/\tau_0 = 10^{15}\,{\rm sec}^{-1}$, then $\tilde\mu \sim 10^8$ Pa, which is about one tenth the yield stress, and means that the individual STZ's are substantially but not excessively more deformable than the system as a whole.  The effective activation energy in the high temperature region, where $\alpha(T) \cong 0$, corresponds to $T_Z + T_E \cong 19,000\,K$, which is roughly consistent with earlier estimates. \cite{MASUHRetal99}
 
\begin{figure}[h]
\centering \includegraphics[height=8 cm]{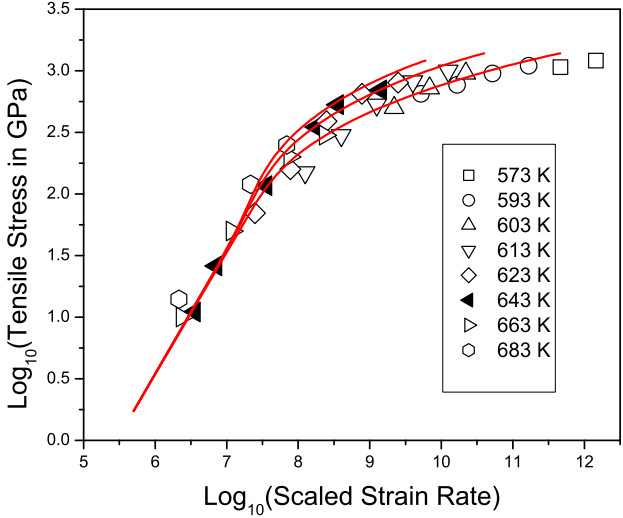}
\caption{\label{BMGstress} (Color online) Tensile stress as a function of the scaled strain rate $2\,\eta_N\,\dot\gamma$. The data points, with temperatures as indicated, are taken from Lu et al. \cite{LUetal03}. The three solid red curves, from bottom to top, are theoretical predictions for temperatures $T = 573\,K$, $643\,K$ and $683\,K$ respectively. }
\end{figure}

Lu et al. \cite{LUetal03} summarize their steady state measurements of flow stress versus strain rate at eight different temperatures by showing that their data nearly collapse, with appreciable scatter, to a single curve of stress as a function of normalized strain rate $2\,\eta_N\,\dot\gamma$, where $\dot\gamma = D_{xx}^{pl}= (2/\sqrt{3})\,\bar D^{pl}$.  Their results are shown in Fig.\ref{BMGstress} along with the STZ predictions for the lowest and highest temperatures used in the experiments, $T=573\,K$ and $683\,K$ respectively, and an intermediate temperature, $T=643\,K$. For computing the theoretical curves in Fig.\ref{BMGstress}, I have used Eq.(\ref{Dpl3}) to compute $\bar D^{pl}$ and the steady-state version of Eq.(\ref{dotchi2}) with $\beta = 1$ to compute $\chi$.  The experimental values of $\dot\gamma$ are all less than or equal to $0.1\,{\rm sec}^{-1}$, therefore $q = \dot\gamma\,\tau_0 \le 10^{-16}$ is completely negligible.  Accordingly, in computing $\chi$, I have used $\hat\chi(q) \approx \hat\chi(0) \equiv \chi_0$.  Then, in addition to using the parameters determined by fitting the viscosity $\eta_N$ as described above, I fit the data in Fig.\ref{BMGstress} by choosing $s_0 = 1.1\,{\rm GPa}$ (the measured room-temperature tensile yield stress divided by $\sqrt{3}$), $s_1 = 1.0\,{\rm GPa}$, $\kappa = 0.5$, and $\chi_0 = 0.6$.  

Having fixed the viscosity parameters, and having chosen not to use a temperature-dependent yield stress or to vary $\beta$, I found strikingly little leeway in determining the remaining parameters $\kappa$ and $\chi_0$.  I did this by fitting the theory to the experimental points for $T = 643\,K$, as shown by the middle curve and the filled triangles in Fig.\ref{BMGstress}. It then becomes clear that some of the apparent scatter in the experimental data is a systematic trend predicted by the theory.  The data collapse is trivial at small strain rates where we are seeing just linear viscosity.  At larger strain rates, the stress as a function of strain rate starts to flatten out as it approaches the yield stress, with -- interestingly -- the stresses at lower temperatures mostly falling below those at higher temperatures. The data go out only to about half the yield stress; therefore the crossover from small-stress to large-stress behavior, which is sensitive to the details of ${\cal R}(\bar s)$, is not explored by the experiments.  

A more sensitive test of the theory is provided by the transient response seen by Lu et al in their constant strain-rate experiments.  To interpret these experiments, we must include the elastic part of the stress-strain relation.  Assuming that the elastic and plastic parts of the rate-of-deformation tensor are simply additive contributions, we can write the $xx$ component of the equation of motion for the stress in the form
\begin{equation}
\label{sigmadot}
{1\over E}\,{d \sigma\over dt} = \dot\gamma - {2\over\sqrt{3}}\,\bar D^{pl}(\bar s,\chi),
\end{equation}
where $E$ is Young's modulus and $\sigma= \sqrt{3}\,\bar s$ is the tensile stress. Then replace time $t$ by $\gamma = \dot\gamma\,t$, and use Eq.(\ref{Dpl3}) to evaluate $\bar D^{pl}$. Equation (\ref{sigmadot}) becomes
\begin{equation}
{d\bar s\over d\gamma}= \tilde E\,\left[1 - {2\,\epsilon_0\over \dot\gamma\,\tau_0}\,e^{-1/\chi}\,{\cal C}(\bar s)\,\Bigl({\cal T}(\bar s)-M(\bar s)\Bigr)\right],
\end{equation}
where, according to data provided in Lu et al., $\tilde E = E/\sqrt{3} = (E/\sigma_y)\,s_0 \cong 50\,s_0$.  This equation for $\bar s(\gamma)$  must be solved along with Eq.(\ref{dotchi2}) for $\chi(\gamma)$, which becomes
\begin{eqnarray}
\label{dotchi3}
\nonumber
{d\chi\over d\gamma}&=& {1\over \tilde c_0\,\dot\gamma\,\tau_0}
e^{-1/\chi}\,\chi\,\Bigl[\Gamma(s)\,\Bigl(1-{\chi\over \chi_0}\Bigr)\cr\\&+&\kappa\,\rho(T)\,\Bigl(1-{T_Z\,\chi\over T}\Bigr)\Bigr].
\end{eqnarray}
At this point, the only free parameters are the dimensionless effective specific heat $\tilde c_0$ and the initial values of $\chi$.  

\begin{figure}[h]
\centering \includegraphics[height=7 cm]{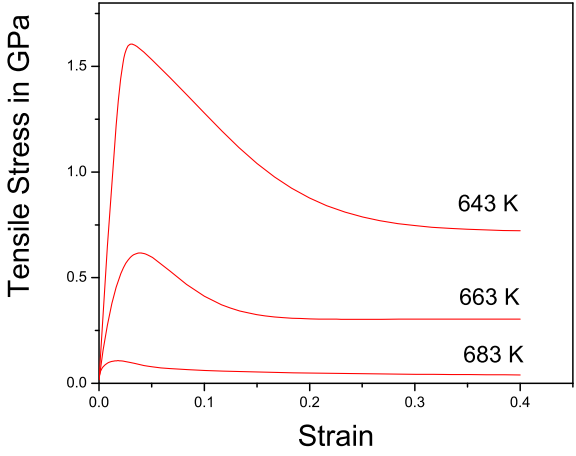}
\caption{\label{Transient1} (Color online) Tensile stress versus strain at three different temperatures as shown.  The strain rate in all three cases is $\dot\gamma = 0.1 {\rm sec.}^{-1}$ The initial effective temperatures, reading from the top curve to the bottom, are $730\,K$, $715\,K$, and $675\,K$ respectively. This figure is to be compared to Fig.1 in \cite{LUetal03}}
\end{figure}

\begin{figure}[h]
\centering \includegraphics[height=7 cm]{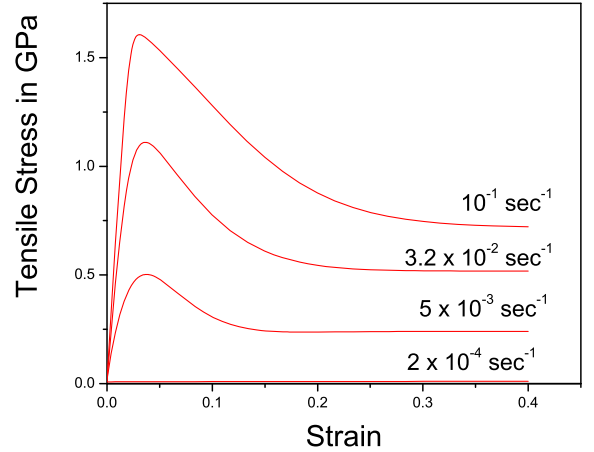}
\caption{\label{Transient2} (Color online) Tensile stress versus strain at three different strain rates as shown. The temperature in all three cases is $T = 643\,K$ The initial effective temperatures $T_i$, reading from the top curve to the bottom, are $730\,K$, $725\,K$, $687\,K$ and $650\,K$ respectively. This figure is to be compared to Fig.2 in \cite{LUetal03}}
\end{figure}

A selection of theoretical stress-strain curves, i.e. tensile stresses $\sigma$ as functions of strain $\gamma$, is shown in Figs.\ref{Transient1} and \ref{Transient2}.  The six separate stress-strain curves shown here, with different temperatures and strain rates, constitute all of the curves shown by Lu et al in their Figs.1 and 2 for which their system is not driven so hard that it fails before deforming plastically.  The curves shown here exhibit characteristic stress peaks at low temperatures and large strain rates.  Both the locations of the peaks at $\gamma \cong 0.05$ and their relaxation to constant flow stresses at about $\gamma \cong 0.2$ are quantitatively consistent with the experimental data.  The relaxation rate is slightly sensitive to the value of $\tilde c_0$, which must be of order unity.  The choice $\tilde c_0 = 0.5$, as used here, seems optimal.  At low temperatures and large strain rates, the theoretical curves exhibit sharp cusps at their peaks, at about the same places where the experimental curves break off, indicating that the sample has failed.  That behavior is illustrated in both figures by the curve for $T = 643\,K$, $\dot\gamma = 0.1$, which is shown here as continuing to where it drops to a flow stress of about $0.5$ GPa, although the experimental curve is not shown in \cite{LUetal03} as continuing much beyond the peak. 

The one adjustment that I have made in computing these stress-strain curves is in the choice of the initial values of the effective temperatures $\chi_i=\chi(\gamma = 0)$.  If the experimental samples were equilibrated at their deformation temperatures, then $\chi_i$ would be equal to $T/T_Z$; but that estimate produces theoretical stress peaks that are higher and sharper than the experimental ones, especially at the lower temperatures and higher strain rates. I can correct these discrepancies without changing the peak positions or the rate at which the system relaxes to the flow stress if I adjust the  $\chi_i$'s to fit the peaks.  My estimates of the $\chi_i$'s, shown in the form of effective temperatures $T_i= \chi_i\,T_Z$, are shown in the figure captions.  Note that these initial effective temperatures remain well below $\chi_0\,T_Z = 960\,K$.  

These experimental samples were quenched to temperatures below the glass temperature, where they had values of $\chi$ equal to the fictive temperature at which their configurational degrees of freedom fell out of equilibrium with the heat bath during the quench.  That fictive temperature seems likely to have been quite high, in most cases larger than the deformation temperature $T$, because these samples could not have been quenched slowly enough to remain in thermal equilibrium very far into the glassy region.  When the samples were then held at temperatures $T$ during rheological measurements, $\chi$ may have moved toward $T$, but would not have reached that temperature except in cases where $T$ was large or the strain rate was very small.  Therefore, I have adjusted the initial effective temperatures as shown. 

It seems to me that the success of the STZ theory in predicting transient behavior is a strong indication that the theory is capturing the underlying physics of dynamic plasticity.  As noted above, essentially all of the system parameters were determined by steady-state measurements, so that there were no numbers that could be adjusted by orders of magnitude in the transient calculations.  That this worked accurately -- despite the fact that the theory relates molecular processes that occur on femtosecond time scales to macroscopically slow plastic deformations -- increases my confidence in the basic features of the theory.  Specifically, this analysis seems to be a stringent test of the assumption that the rates of change of STZ populations are determined by the rate at which disorder is generated during deformation, and the accompanying idea that disorder in amorphous materials is accurately described by an effective temperature.

\section{Concluding remarks}
\label{conclusions}

I close by stating just a few brief opinions about open questions and further directions for research.  

As implied in the text, I think that the least well developed element of the theory presented in this paper is the derivation of the thermal entropy-production term, $(d\,S_c/d\,t)_{therm}$, given in Eq.(\ref{Sctherm}).  This term seems likely to be strongly model dependent because it involves more than just the STZ degrees of freedom.  It also may control some especially interesting physical properties such as shear-band formation.  

Perhaps the most fundamental theoretical challenge is to understand the limits of validity of the STZ theory.  How and when does it break down at high deformation rates? Or during large excursions from steady state behavior where the separation of time scales described in Sec. \ref{STZequations} becomes invalid?  We know that, at some point, strongly driven amorphous materials must change from deforming slowly like solids to flowing rapidly like liquids.  Where and how does this happen?  Manning and I may have seen a clue in \cite{JSL-MANNING07}, where it appears that the effective temperature diverges at a large but finite strain rate of order $\tau_0^{-1}$.  This line of investigation has been opened by the work of Haxton and Liu \cite{HAXTON-LIU07}.  More simulations and experiments along these lines should be very interesting.   

Looking from a broader point of view, I have argued elsewhere \cite{JSL-Handbook05} that a principal goal of research in  solid mechanics ought to be to bridge the gap between atomistic physics and engineering practice.  My prime example of how far behind we are in this area is fracture mechanics.  It is well known that advancing cracks undergo instabilities that are qualitatively similar to sidebranching instabilities in dendritic crystal growth.  The latter instabilities have been well understood for about half a century, and that insight has provided the basis for major advances in solidification processing.  As yet, we have no comparable understanding of the analogous instabilities in fracture. In \cite{JSL-frac00}, I suggested a way in which plasticity theory might be brought to bear on this problem; but, at that time, the STZ theory was not well enough developed for much progress to be made.  We are now beginning, in \cite{MLC07} for example, to understand in an STZ context how driven amorphous materials may become unstable against spatially nonuniform deformations such as shear bands.  Reference \cite{BLLP07} is explicitly an attempt to move in the direction of ordinary fracture.  The STZ theory ought to be relevant at least to slow, ductile failure in either shearing or tensile modes.  Can it also predict the behavior of fast brittle cracks? 

\begin{acknowledgments}
This research was supported by DOE grant number DE-FG03-99ER45762.  I thank Lisa Manning for her essential  help and encouragement in this project; and I thank Eran Bouchbinder, Michael Falk, and Itamar Procaccia for their critical readings of many versions of this paper.  
\end{acknowledgments}

\end{document}